\pdfoutput=1
\documentclass[aps,pra,preprint,tightenlines,showpacs,showkeys,superscriptaddress]{revtex4-1}
\usepackage{graphicx}
\usepackage{amssymb}
\usepackage[squaren, cdot]{SIunits}
\usepackage[latin9]{inputenc}

\begin{document}
\title[Fast transport, atom sample splitting, and single-atom qubit supply ...]{Fast transport, atom sample splitting, and single-atom qubit supply in two-dimensional arrays of optical microtraps\\
---\\
Reference: New J. Phys. \bf{14}, 123034 (2012)\\
--- 
}
\author{Malte Schlosser}
\author{Jens Kruse}
\author{Christian Gierl}
\author{Stephan Teichmann}
\author{Sascha Tichelmann}
\author{Gerhard Birkl}
\email[]{gerhard.birkl@physik.tu-darmstadt.de}
\affiliation{Institut f\"{u}r Angewandte Physik, Technische Universit\"{a}t Darmstadt, Schlossgartenstra\ss e 7, 64289 Darmstadt, Germany}
\date{\today}

\begin{abstract}
Two-dimensional arrays of optical micro-traps created by microoptical elements present a versatile and scalable architecture for neutral atom quantum information processing, quantum simulation, and the manipulation of ultra-cold quantum gases. In this article, we demonstrate advanced capabilities of this approach by introducing novel techniques and functionalities as well as the combined operation of previously separately implemented functions. We introduce piezo-actuator based transport of atom ensembles over distances of more than one trap separation, examine the capabilities of rapid atom transport provided by acousto-optical beam steering, and analyze the adiabaticity limit for atom transport in these configurations. We implement a spatial light modulator with 8-bit transmission control for the per-site adjustment of the trap depth and the number of atoms loaded. We combine single-site addressing, trap depth control, and atom transport in one configuration for demonstrating the splitting of atom ensembles with variable ratio at predefined register sites. Finally, we use controlled sub-poissonian preparation of single trapped atoms from such an ensemble to show that our approach allows for the implementation of a continuous supply of single-atom qubits with high fidelity. These novel implementations and their combined operation significantly extend available techniques for the dynamical and reconfigurable manipulation of ultracold atoms in dipole traps.
\end{abstract}
\maketitle

\section{Introduction}
\label{sec:intro}
\begin{figure}
\begin{center}
 \includegraphics[width=0.75\linewidth]{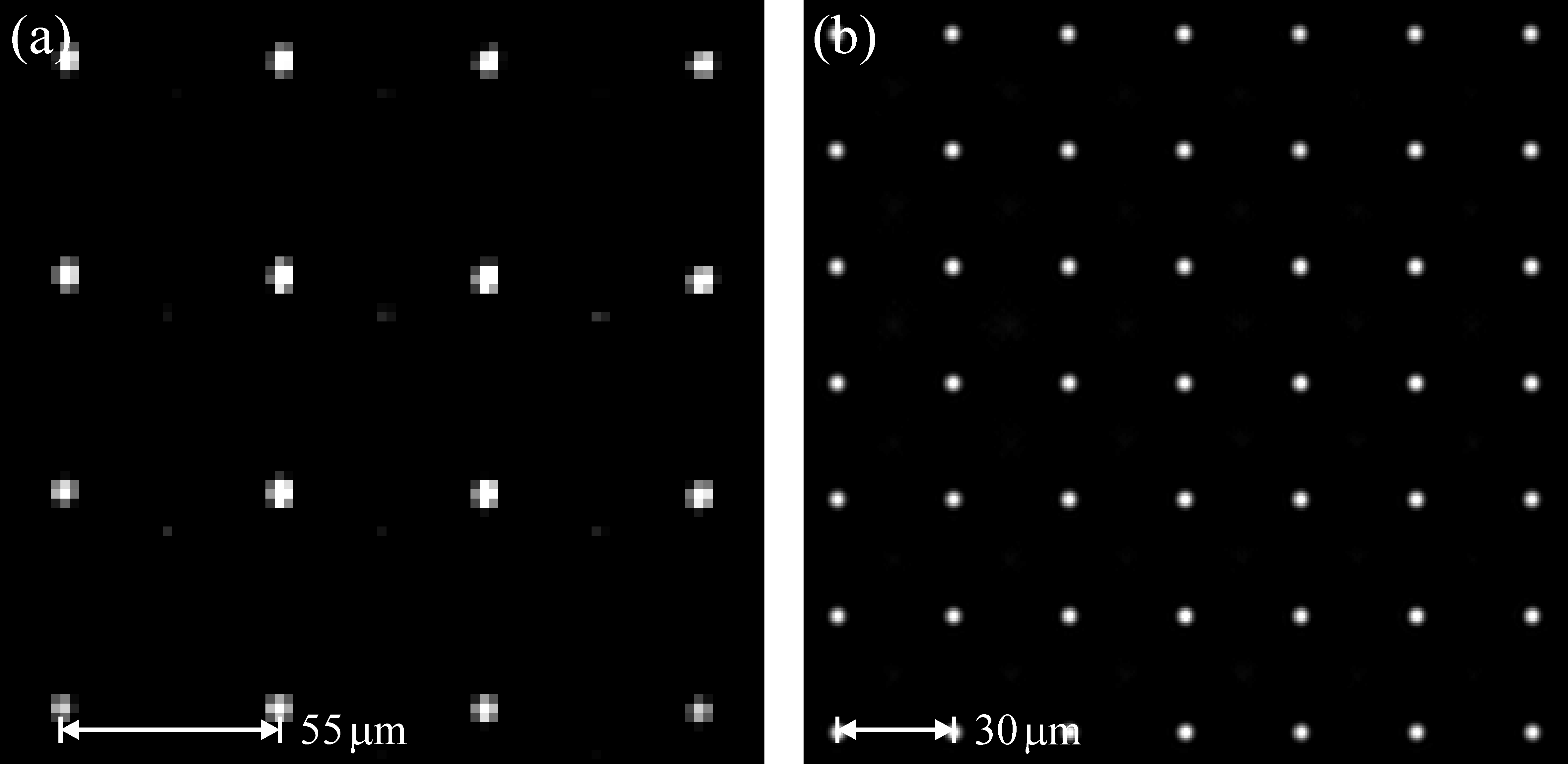}
 \end{center}
\caption{
Two-dimensional spot patterns generated by two different microlens arrays. Each focal point represents a dipole trap for neutral atoms. (a) Dipole trap array of $\unit{55}{\micro\meter}$ pitch, created by demagnifying the focal plane of a $\unit{125}{\micro\meter}$-pitch microlens array; (b) focal plane of a $\unit{30}{\micro\meter}$-pitch lens array used without further demagnification. 
}
\label{fig:mla}
\end{figure}

In the quest of a physical system suitable for the successful implementation of quantum information processing (QIP) \cite{2000:QCQ:544199}, a series of candidates have been studied ranging from solid state physics to quantum optics \cite{Bouwmeester:2001,Beth:2005:QIP:1076259,Everitt:2005:EAQ:1205130,Schleich:2007:EQI:1526289}.
During the last decade, several of these approaches have experienced remarkable progress towards fulfilling the essential requirements for quantum computation as listed for example in \cite{2000ForPh..48..771D}.
A subset of these candidates relies on trapped neutral atoms (atom ensembles and single atoms) as intrinsically identical systems which feature decoupling from the environment to a high degree and an extraordinary extent of control of their internal and external degrees of freedom (for an overview see \cite{QIP.special.issue}).\\
Thus, quantum information processing with neu\-tral a\-toms but also the investigation of ultra-cold atomic quantum gases in external confining potentials (see \cite{BlochRMP2008} and references therein) strongly profit from versatile optical trapping configurations. For this purpose, we have introduced the application of micro-fabricated optical elements for atom optics and QIP with atoms \cite{2001OptComm,Buchkremer:2001} and have implemented versatile architectures providing scalability, reconfigurability, and state-of-the-art technology \cite{PhysRevLett.89.097903,2007ApPhB..86..377L,PhysRevA.81.060308,PhysRevLett.105.170502,springerlink:10.1007/s11128-011-0297-z}, laying the foundation for a comprehensive system which is capable of incorporating all respective requirements for QIP \cite{2000ForPh..48..771D}.
In concert with other implementations based on optical trapping configurations, experimental progress has been reported for example for the site-selective manipulation of spins and thus for the required one-qubit gates \cite{PhysRevA.81.060308,PhysRevLett.93.150501,PhysRevLett.103.080404,MEMS_Addressing,2011Natur.471.319W}, for the reliable coherent storage and transport of atomic quantum states \cite{PhysRevLett.105.170502,PhysRevLett.91.213002,2007NatPh...3..696B}, and for the demonstration of state-selective detection and of near-deterministic preparation of single atoms per site \cite{springerlink:10.1007/s11128-011-0297-z,SchlosserSingleAtoms,Greiner2002,2001Natur.411.1024S,2007NatPh...3..556N,Mott_Insulator,2009:Greiner:Microscope,2011NatPh...6..951G}.\\
The combination of coherent atom transport, as demonstrated in \cite{PhysRevLett.105.170502} and extended here, and of trap configurations with optimized dimensions also provides a clear path towards the implementation of two-qubit gates \cite{2003NatMandel,2007Natur.448..452A,PhysRevLett.104.010502,PhysRevLett.104.010503} being the only missing link towards realizing all elements of a functional quantum processor in our approach. As discussed in \cite{springerlink:10.1007/s11128-011-0297-z}, our system is well suited for extending the work on Rydberg-atom based gate operations \cite{PhysRevLett.104.010502,PhysRevLett.104.010503} to the multi-qubit regime in a scalable two-dimensional architecture.\\
Key elements of our approach are two-dimensional (2D) registers of optical micro-potentials created by microlens arrays (see Sec.\ \ref{sec:array}) in which stored $^{85}$Rb atoms serve as carriers of quantum information. We obtain a system of well resolved qubit register sites with each site being defined by the focal spot of an individual microlens out of the 2D lens array (Fig.\ \ref{fig:mla}).
This has two significant consequences:\\
\begin{enumerate}
\item The position of the register sites is linked to the position of the microlens array and to the
incident angle of the trapping laser beam. We have implemented techniques for position control and realized the coherent transport of atomic quantum states in a scalable shift register \cite{PhysRevLett.105.170502}. 
\item The light illuminating each particular microlens can be controlled by a spatial light modulator (SLM). This allows us to initialize, manipulate, and probe the stored qubits in a reconfigurable fashion. As a direct result, qubits can be coherently addressed either at all register sites in parallel \cite{2007ApPhB..86..377L} or in a site-selective fashion \cite{PhysRevA.81.060308}.\\
\end{enumerate}
\newpage
{\bf Advances presented in this publication}\\

In this article, we present a series of significant extensions of our previous work. These aim towards overcoming potential limitations of our previous implementations (e.g.\ increased speed of atom transport), experimentally demonstrate extended functionalities by combining previously separated functions in one combined configuration, or add completetly new capabilities not demonstrated before (e.g.\ for continuous supply of single-atom qubits):\\
%
%
\begin{enumerate}
\item We implement a new generation of micro-fabricated lens arrays with reduced dimensions in Sec.\ \ref{sec:array}. The novel elements enable the creation of microtrap arrays with the number of traps increased to $10^4$.
\item For faster qubit transport, we examine in Sec.\ \ref{sec:aom} an implementation of atom transport based on acousto-optical beam steering which is capable of pushing time constants for transport from the millisecond to the microsecond regime.
\item We introduce piezo-actuator enabled positioning and demonstrate atom transport over a full trap pitch with this technique in Sec.\ \ref{sec:shift}. This method allows for atom transport without the potentially degrading effects of skewed illumination of microlenses.
\item In Sec.\ \ref{sec:adiabatic}, we discuss the adiabaticity limit for atom transport in dipole trap arrays and show that many transport cycles are possible within the coherence time of our configuration.
\item We demonstrate in Sec.\ \ref{sec:lcd} that an analog control of the depth of every single microtrap within an array can be achieved allowing for the adjustments of the trap parameters over the full array and for an individual control of the number of atoms loaded in each trap.
\item In Sec.\ \ref{sec:combined}, we demonstrate the implementation of a combined system for atom transport and site-selective manipulation allowing for the preparation, transport, and splitting of atom ensembles in reconfigurable trap patterns. With this technique, one set of register sites can e.g.\ serve as a set of reservoirs for the repeated extraction of small atom samples in complex architectures for QIP.
\item In Sec.\ \ref{sec:singleatom}, we show that this system can be extended to a configuration for a continuous supply of {\it single-atom} qubits: we experimentally demonstrate that by adding the controlled induction of two-body collisions, we can convert an initially poissonian number distribution of trapped atoms (e.g.\ present in the extracted traps of Sec.\ \ref{sec:combined}) to a sub-poissonian atom number distribution with only 0 atom or 1 atom per trap and at least 50 \% probability of having exactly 1 atom per trap. We can determine the atom number with high reliability thus demonstrating the key elements for a fully deterministic supply of single-atom qubits for a scalable quantum computation architecture.\\
\end{enumerate}
\section{Qubit registers based on atoms in arrays of optical micro-potentials}
\label{sec:array}%
\begin{figure}
\begin{center}
\includegraphics[width=0.75\linewidth]{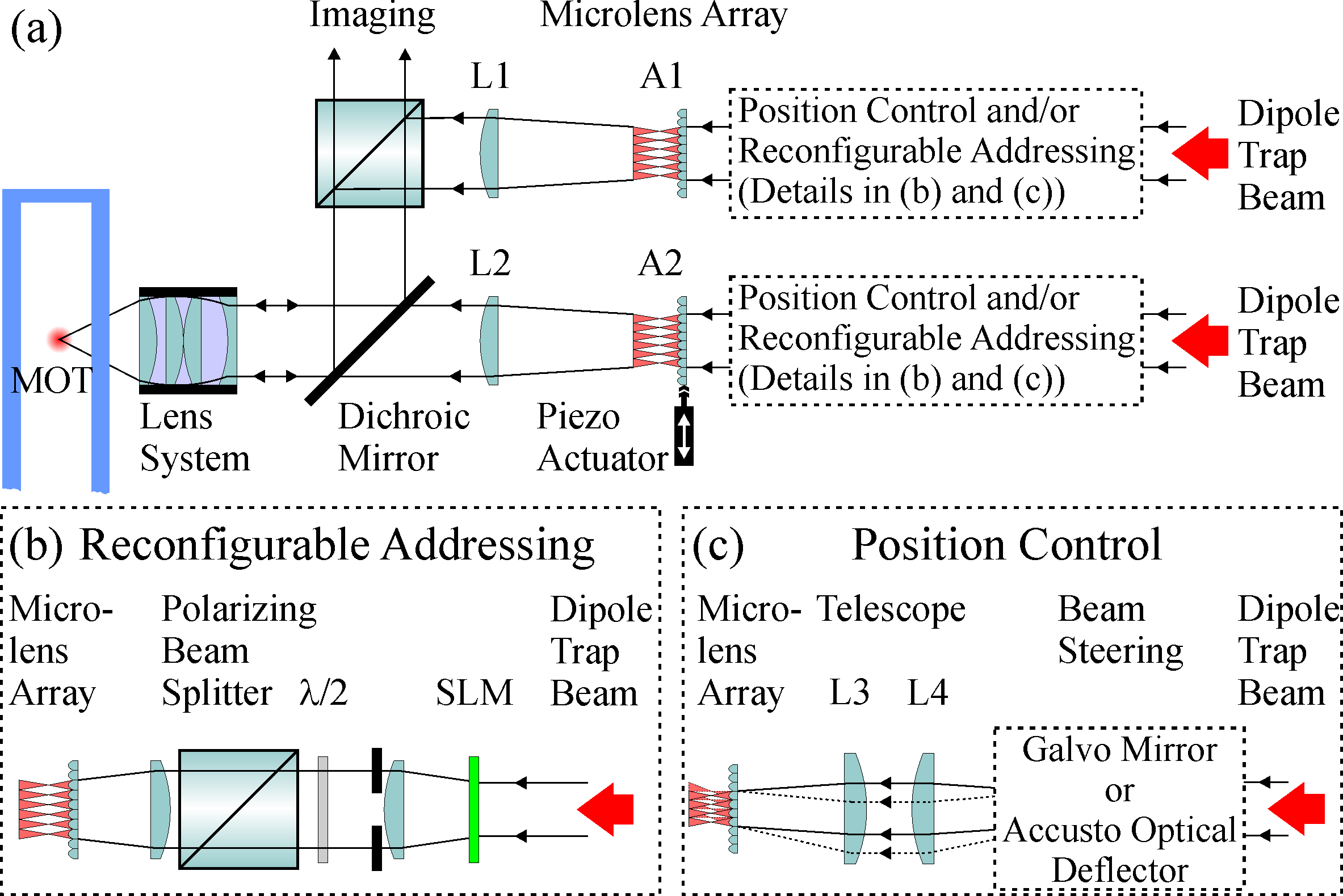}
\end{center}
\caption{
(color online). Schematic view of the experimental setup. (a) The microlens arrays A1 and A2 are illuminated by two trapping laser beams. The resulting spot patterns are combined at a dichroic mirror and re-imaged into the vacuum chamber. The incident dipole trap beams can be position-controlled and site-selectively addressed. (b) A spatial light modulator (SLM) is used to control the light power addressing each microlens. (c) A galvo mirror or an acousto-optical beam deflector can be used to control the incident angle of the trapping laser beam on the microlens array A1 and/or A2 and therefore the position of the microtraps in the focal plane. Position control can be implemented through a piezo actuator in addition (shown for A2 in (a)).
}
\label{fig:setup}
\end{figure}

The use of micro-fabricated optical elements takes atom optics to the micro-regime, where a single conventional setup can be extended to thousands of parallelized implementations \cite{2001OptComm,Buchkremer:2001,2007BirklLasphotrev}. In the case of a focused beam dipole trap, employing an array of microlenses leads to the creation of a 2D dipole trap array. Figure \ref{fig:mla} shows details of the light fields generated by the microlens arrays used for the experiments presented in the following sections. The spot pattern of Fig.\ \ref{fig:mla}(a) results from the demagnified focal plane of a quadratic-grid microlens array of $\unit{125}{\micro\meter}$ pitch, consisting of $40\times 40$ spherical lenses of $\unit{100}{\micro\meter}$ diameter and a numerical aperture $NA = 0.05$.
The novel element creating the dipole traps of Fig.\ \ref{fig:mla}(b) contains $166\times 166$ lenses with a pitch of $\unit{30}{\micro\meter}$, a diameter of $\unit{26}{\micro\meter}$, and a numerical aperture $NA = 0.144$.
Due to limitations in laser power, in the experiments presented in this work we typically implement about 50 trapping sites loaded with atoms. Fully exploiting the available number of the second generation microlenses will result in an array of microtraps with well over $10^4$ trapping sites.\\
The experimental setup is shown in Fig.\ \ref{fig:setup}. It features two independent beam paths for two separate microlens arrays. The trapping laser beams address subsets of the particular microlens array with light at a wavelength of $\unit{782.7}{\nano\meter}$ (array A1) and of $\unit{795.8}{\nano\meter}$ (array A2). With a 2D spatial light modulator 
based on a liquid crystal display (LCD) we can control the light power illuminating each microlens (Fig.\ \ref{fig:setup}(b)) \cite{PhysRevA.81.060308}.
The SLM can be inserted either in the beam path of A1 or A2. Additionally, the microtrap arrays are position controlled either by the use of a piezo actuator (shown for A2 in Fig.\ \ref{fig:setup}(a)) or a variation of the trapping lasers' incident angle (Fig.\ \ref{fig:setup}(c)) provided by acousto-optics (Sec.\ \ref{sec:aom}) or a galvo mirror \cite{PhysRevLett.105.170502}.
The beam paths originating from both arrays are combined at a dichroic mirror.
The arrays of laser spots are re-imaged into the vacuum chamber employing an achromatic lens (L1, L2) and a lens system of $NA=0.29$ and focal length of $\unit{35.5}{\milli\meter}$. The re-imaging provides flexibility regarding the characteristics of the implemented trap array, since it allows us to rapidly exchange the micro-optics and to demagnify the potential geometries to smaller dimensions if required.\\
\newpage
In Sec.\ \ref{sec:shift} we use a trap array as shown in Fig.\ \ref{fig:mla}(b) with a pitch of $\unit{30}{\micro\meter}$, a trap waist of $w_0=\unit{2.5}{\micro\meter}$, and a typical trap depth of $U_0=k_B\times\unit{1}{\milli\kelvin}$ without further demagnification. The corresponding trap frequencies are $\omega_r=\sqrt{4U_0/(mw_0^2)}=2\pi\times\unit{39.8}{\kilo\hertz}$ for the radial and $\omega_z=\sqrt{2 U_0/(mz_R^2)}=2\pi\times\unit{2.9}{\kilo\hertz}$ in the axial direction with the Rayleigh range $z_R=\pi w_0^2/\lambda=\unit{25}{\micro\meter}$.
For the experiments of Sec.\ \ref{sec:lcd} and Sec.\ \ref{sec:combined} we use a $\unit{55}{\micro\meter}$-pitch register (Fig.\ \ref{fig:mla}(a)) with a trap waist of $\unit{3.8}{\micro\meter}$, Rayleigh range $z_R=\unit{57}{\micro\meter}$, a typical depth of $k_B\times\unit{0.1}{\milli\kelvin}$ and frequencies of $\omega_r=2\pi\times\unit{8.3}{\kilo\hertz}$ and $\omega_z=2\pi\times\unit{0.4}{\kilo\hertz}$ created by demagnifying the focal plane of a $\unit{125}{\micro\meter}$-pitch microlens array. Unless otherwise noted, all of the experiments presented here are performed with small ensembles of $^{85}$Rb atoms (atom number per site $n\approx 10$) captured in a magneto-optical trap (MOT) and loaded into the superimposed microtrap arrays during a sequence of optical molasses. Typical lifetimes in this setup are on the order of one second and coherence times $T_2'$, limited by homogeneous dephasing and spontaneous scattering, on the order of $\unit{70}{\milli\second}$ have been observed \cite{PhysRevLett.105.170502}. The stored atom ensembles are detected employing fluorescence imaging with per-site resolution, where the high-NA lens system used for re-imaging of the laser spots also serves to collect the fluorescence of the trapped atoms.
%
%
\section{Shifting micro-potentials with acousto-optics}
\label{sec:aom}
%
\begin{figure}
\begin{center}
 \includegraphics[width=0.75\linewidth]{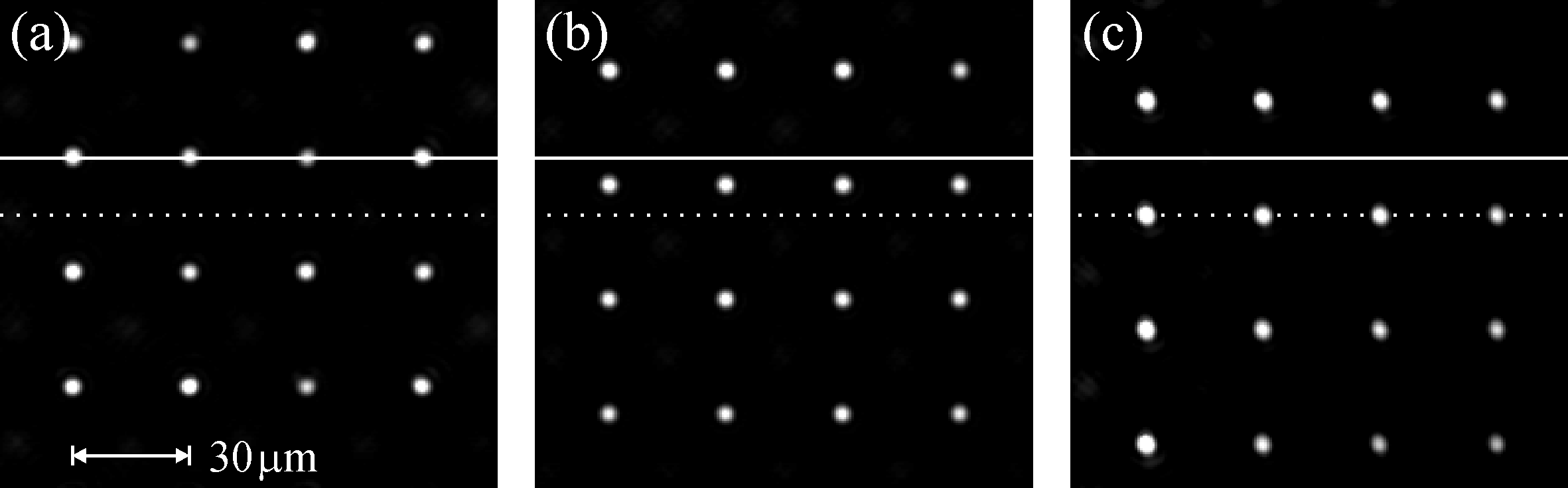}
 \end{center}
\caption{
Spot pattern in the focal plane of a $\unit{30}{\micro\meter}$-pitch microlens array. An acousto-optical deflector is used to incline the trapping laser beam on the lens array which causes the foci to shift laterally. (a) Initial position; (b) shift by a quarter and (c) by half of a trap separation of $\unit{30}{\micro\meter}$.
}
\label{fig:aom}
\end{figure}
%
In architectures based on micro-optics, there are several possibilities to manipulate the position of trapped atoms (see also Sec.\ \ref{sec:shift}). One is the variation of the incident angle of the trapping laser beam on the microlens array (Fig.\ \ref{fig:setup}(c)) which causes the foci to shift laterally within the focal plane. For a microlens array with a pitch of $\unit{125}{\micro\meter}$ ($\unit{30}{\micro\meter}$) and NA = 0.05 (0.144), a deviation from the normal angle of $\unit{\pm 3.6}{\degree}$ ($\unit{\pm 8.2}{\degree}$) results in the shift over a full trap separation. Based on this technique, we have previously implemented a multi-step quantum shift register and demonstrated the coherent transport of atomic quantum states between adjacent trapping sites \cite{PhysRevLett.105.170502}. In this realization, the trapping laser is deflected by a galvo mirror, where the pivot point is imaged onto the microlens array using a telescope with unity magnification $V=1$ (L3, L4 in Fig.\ \ref{fig:setup} (c) with a focal length of $f=\unit{200}{mm}$). This method provides the capability of dynamic modification of the incident angle with sufficient magnitude on a millisecond timescale, reaching the limit in scan speed of the galvo mirror.\\
A significant speed-up of atom transport in a microtrap register can be expected from the implementation of acousto-optics for beam steering. In an advanced set of experiments, we have replaced the galvo mirror by an acousto-optical deflector (AOD) and have changed the telescope to $V=0.4$ by inserting a $f=\unit{100}{\milli\meter}$ lens at position L4 and a $f=\unit{40}{\milli\meter}$ lens at position L3. In this configuration, the AOD covers a range of $\unit{14.8}{\degree}$. We analyzed its capabilities for position control by observing the focal plane of a $\unit{30}{\micro\meter}$-pitch array at position A2. In Fig.\ \ref{fig:aom}(a) the focal spots are shown at their initial position with the incident trapping laser beam under normal angle. The images of Fig.\ \ref{fig:aom}(b) and \ref{fig:aom}(c) display the shifted focal plane for a deviation from the normal angle of $\unit{4.1}{\degree}$ and $\unit{8.2}{\degree}$, resembling a shift over one quarter and one half of the trap separation, respectively.
The accessible range of deflection is already sufficient for the implementation of rapid acousto-optic enabled atom transport in the $\unit{55}{\micro\meter}$ trap register with required deflection angle variation of $\unit{7.2}{\degree}$, while a shift over a full trap separation in the $\unit{30}{\micro\meter}$ trap register corresponds to a variation of the deflection angle of $\unit{16.4}{\degree}$ which would require slight modifications of the optical setup ($V = 0.36$).
Since for AODs beam steering is accomplished at least one order-of-magnitude faster in comparison to galvo mirrors, technical limitations are lifted and the accessible transport times are limited by adiabaticity requirements for the atom transport as discussed in Sec.\ \ref{sec:adiabatic}.
\section{Atom transport using a piezo actuator}
\label{sec:shift}
\begin{figure}
\begin{center}
 \includegraphics[width=0.75\linewidth]{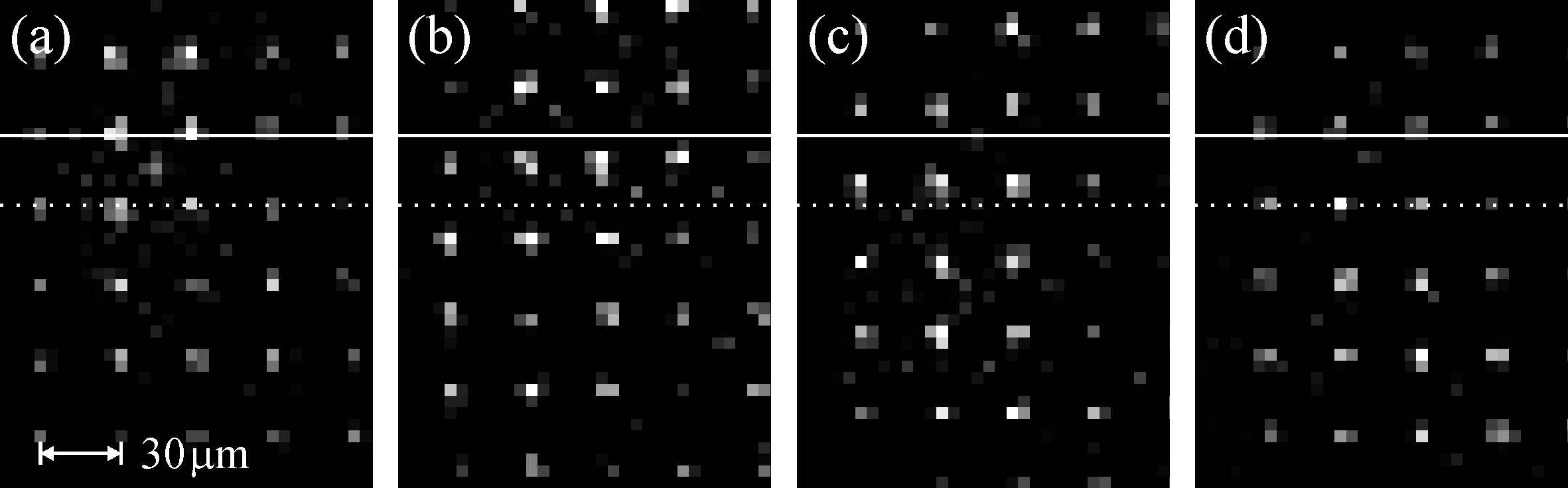}
 \end{center}
\caption{
(a) Atoms stored and transported in a $\unit{30}{\micro\meter}$-pitch register. (b,c) A piezo actuator is used to shift the microlens arrays' position with the atoms being transported along. (d) Position of maximum vertical displacement of $\unit{30.6}{\micro\meter}$. Images are averaged 10 times.
}
\label{fig:piezo}
\end{figure}

Transport mechanisms relying on the skewed irradiation of optical elements introduce optical aberrations of the resulting potentials. This puts limits on the achievable distance with respect to the initial position in a single operation, an obstacle that we have been able to overcome with the repeated handover of qubits between neighbouring trapping sites in a shift register operation as demonstrated in \cite{PhysRevLett.105.170502}.\\
In contrast, the approach of piezo-actuator based atom positioning and transport is completely free of additional optical aberrations. Here, the incident angle of the trapping laser is kept constant since the microlens array itself is mounted to a piezo-controlled positioning system and moved to the desired position (see piezo actuator mounted to array A2 in Fig.\ \ref{fig:setup}(a)). This causes the focal spots to shift accordingly and the trapped atoms to be transported along. The experimental implementation of this method 
for an array with $\unit{30}{\micro\meter}$ pitch is shown in Fig.\ \ref{fig:piezo}. Atoms are loaded into the dipole trap array and subsequently transported over variable distances. Images of the atom ensembles stored in the dipole trap array are taken for zero piezo voltage (Fig.\ \ref{fig:piezo}(a)), for two intermediate voltages (Fig.\ \ref{fig:piezo}(b,c)) and for the maximum applied voltage inducing a shift of $\unit{30.6}{\micro\meter}$ (Fig.\ \ref{fig:piezo}(d)). Technical limitations for the speed of the atom transport arise from mechanical resonances with frequencies on the order of $\unit{40}{\kilo\hertz}$. This suggests that this technique can be extended to transport speeds with resulting time\-scales well below $\unit{1}{ms}$ for connecting neighbouring trapping sites, reaching the boundaries implied by the need for adiabaticity (see Sec.\ \ref{sec:adiabatic}). In addition, piezo controlled positioning systems with travel ranges of several hundred microns are available. Upgrading our setup with a 2D device of $\unit{300}{\micro\meter}$ range in future implementations will grant the ability to shuffle quantum information between $10 \times 10$ register sites in a single-step operation.
%
%
\section{Adiabaticity of qubit transport}
\label{sec:adiabatic}
The ability to transport quantum information in an unperturbed fashion is an essential ingredient in large scale quantum processing architectures. It allows not only the development of concepts with spatially separated functional subsections \cite{springerlink:10.1007/s11128-011-0297-z,PhysRevA.84.032322}, e.g.\ for quantum state preparation, storage, and processing, but also to apply complex algorithms involving qubits initially located at well separated positions. In addition, position control is essential for the feasibility of two-qubit gates, since distance is a critical parameter when interaction becomes indispensable.\\
As a matter of principle, the quantum state of a transported qubit has to be preserved. Depending on whether the qubit is encoded in internal or external degrees of freedom, different constraints arise. Although a more severe limitation can be expected for external-state qu\-bits, also for internal-state qu\-bits the increase of vibrational quanta has to be suppressed to a high degree during qubit transport, since the strength of interaction in a two-qubit gate operation in most cases is dependent on the spatial wavefunction.
This criterion puts constraints on the functional form $x(t)$ of the transport trajectory and fundamentally limits the achievable minimum transport time $T_{trans}^{min}$.\\
There has been extensive work on the analysis of error (heating) sources and optimization of transport processes in micro-structured ion traps 
\cite{PROP:PROP200610324,PROP:PROP200610326,HuculYOMHR08}.
%
\begin{figure}
\begin{center}
 \includegraphics[width=0.75\linewidth]{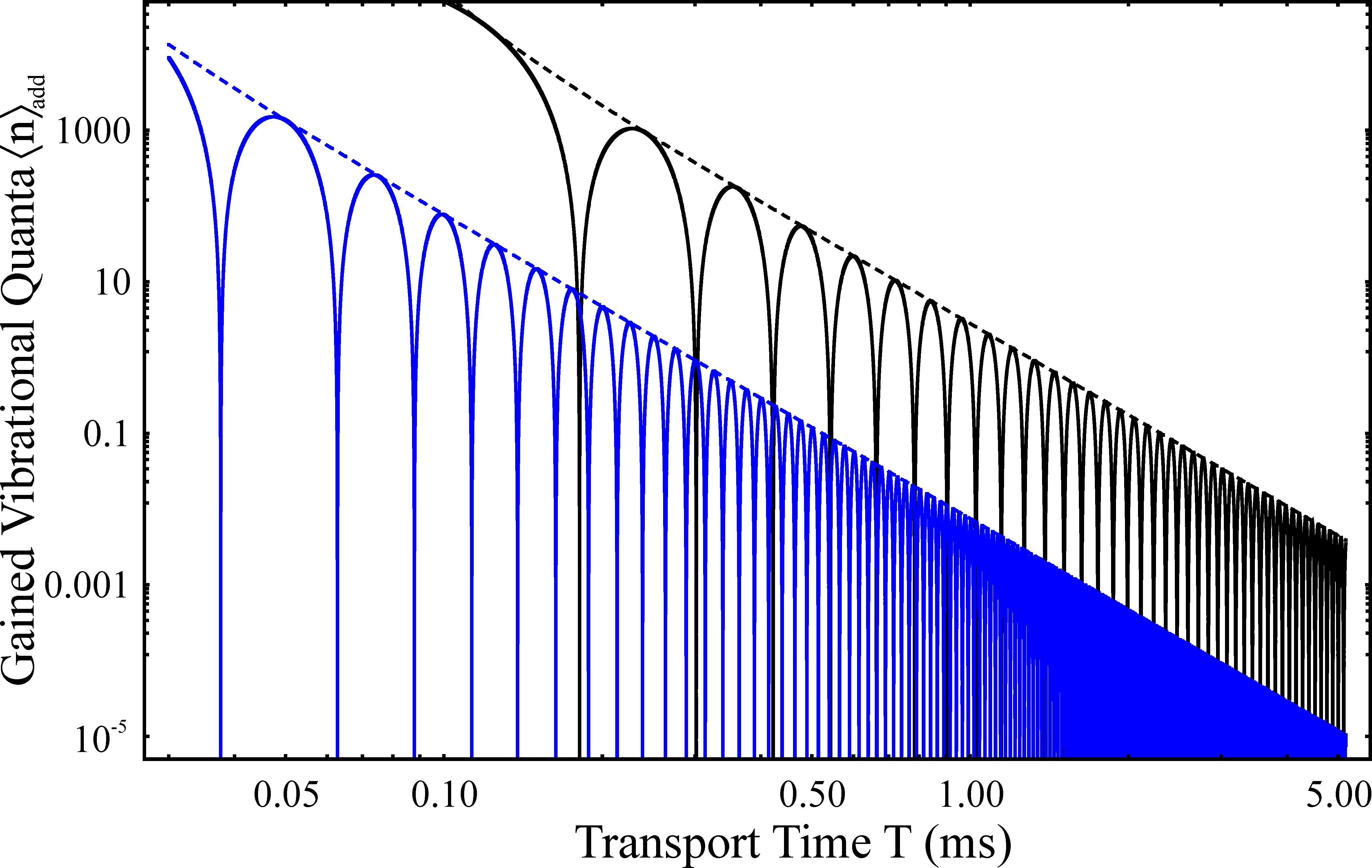}
 \end{center}
\caption{
(color online). Average number of gained vibrational quanta $\langle n(T)\rangle_{add}$ during transport over one trap separation as a function of transport time $T$ for the $\unit{55}{\micro\meter}$-pitch array (black, solid) and the $\unit{30}{\micro\meter}$-pitch array (blue, solid).
The oscillatory function $\langle n(T)\rangle_{add}$ becomes zero once every oscillation period and an upper bound for the gained motional energy is given by its envelope (dashed line). In the $\unit{30}{\micro\meter}$-pitch array heating is suppressed by two orders of magnitude for a given transport time and transport over one trap separation can be achieved four times faster for a given limit of gained energy with respect to the $\unit{55}{\micro\meter}$-pitch array.}
\label{fig:TheoTransport}
\end{figure}
%
The results hold for harmonically bound particles in general, such as optically trapped atoms with a kinetic energy much smaller than the trap depth and small excursions of the wavepacket from the trap center. In a forced parametric oscillator with
\begin{equation}\label{eqn:osc}
\mathcal{H}(t)=\frac{p^2}{2m}+\frac{m\omega^2(t)}{2}q^2+m\ddot x(t)q
\end{equation}
having momentum p and deviation from the trap center q, variations in the trap frequency $\omega(t)$ and an acceleration $\ddot x(t)$ due to transport are the dominant heating sources as they induce transitions between vibrational eigenstates. Here, the impact of both aspects is largely separable and the amount of motional energy which is gained during transport is solely determined by the respective classical quantities \cite{PROP:PROP200610326,HuculYOMHR08}.
Parametric heating occurs if the width of the wavepacket can not follow a variation of $\omega(t)$ adiabatically and is commonly accommodated with the adiabaticity constraint $\dot{\omega}\ll\omega^2$. For the transport in optical dipole traps a variation of the trapping frequency is mainly caused by optical aberrations of the trapping potential and can be suppressed by a carefully designed optical system. In addition, even for nonideal realizations, the adiabaticity criterion for a variation in $\omega(t)$ is easily fulfilled for any reasonable transport parameters \cite{HuculYOMHR08}.\\
Thus, the limiting factor is the nonvanishing displacement after transport time $T$ of the wavepacket from the trap center due to inertial forces during transport. The average number of transferred vibrational quanta is given by \cite{PROP:PROP200610326,HuculYOMHR08}
\begin{equation}\label{eqn:nadd}
\langle n(T)\rangle_{add}=\frac{mS^2\pi^4\omega_0\cos^2\left(\omega_0 T/2\right)}{\hbar\left(\pi^2-\omega_0^2 T^2\right)^2}
\end{equation}
assuming a sinusoidal transport function $x(t)$, a fixed vibrational frequency $\omega_0$ and a transport distance $S$. Figure \ref{fig:TheoTransport} displays a graph of $\langle n(T)\rangle_{add}$ as a function of the transport time $T$ for the two microtrap arrays with pitch of $\unit{30}{\micro\meter}$ (blue) and $\unit{55}{\micro\meter}$ (black) and experimental parameters of Sec.\ \ref{sec:array}.\\
In the adiabatic case which we discuss first $\langle n(T)\rangle_{add}$ is given by the envelope of Eqn. \ref{eqn:nadd} setting $\cos^2\left(\omega_0 T/2\right)=1$ (dashed lines in Fig.\ \ref{fig:TheoTransport}). Here, heating is suppressed by two orders of magnitude for the novel $\unit{30}{\micro\meter}$-pitch array with respect to the $\unit{55}{\micro\meter}$-pitch array for any given transport time and a travel distance of one full trap separation. In the case of a fixed limit of transferred motional energy transport can be achieved four times faster in the novel array. If we restrict the number of gained vibrational quanta to $\langle n(T)\rangle_{add}\leq 1$ we obtain $T_{trans}^{min}(\unit{55}{\micro\meter}) = \unit{1.3}{\milli\second}$ and $T_{trans}^{min}(\unit{30}{\micro\meter}) = \unit{294}{\micro\second}$ for the two microtrap arrays and a transport over the respective pitch. Therefore, qubit transport in our second generation microtrap architecture with $\unit{30}{\micro\meter}$ pitch clearly profits from the reduced transport time of AOD and piezo actuator configurations. Even if we apply the criterion $\langle n(T)\rangle_{add}\leq 1$, strictly valid for external-state qubits, to our internal-state qubit implementation \cite{2007ApPhB..86..377L} which should be less sensitive, already about 200 shift operations with a travel distance of one pitch are possible during the coherence time in the adiabatic limit.\\
Furthermore, making use of the oscillatory behaviour of Eqn. \ref{eqn:nadd} where $\langle n(T)\rangle_{add}=0$ for transport times being multiples of the oscillation period of the atom in the trap (Fig.\ \ref{fig:TheoTransport}), qubit transport on a timescale of single oscillation periods is an evident objective and subject to active research \cite{0295-5075-83-1-13001,chen:134103,1367-2630-13-11-113017}. Elaborating this idea leads to advanced non-adiabatic heating-free transport protocols developed in the framework of quantum optimal control \cite{PROP:PROP200610324,PhysRevA.79.020301,PhysRevA.84.043415,Negretti:12} with a potential gain in transport speed of one order of magnitude and the potential to perform thousands of shift operations within the coherence time.
\section{Reconfigurable site-selective control of trap depth}
\label{sec:lcd}
The ability to selectively address every single qubit is one of the main requirements for quantum information processing \cite{2000ForPh..48..771D}. In our setup of 2D patterns of microtraps generated by 2D microlens arrays, single-site addressability is an inherent feature, since each trapping site is interconnected to the light addressing the corresponding microlens. Benefitting from this, we have introduced an implementation employing an SLM for the site-selective switching of each lenslet in an array of microlenses and demonstrated the coherent manipulation of the quantum state of stored qubits as well as the creation of reconfigurable trap patterns \cite{PhysRevA.81.060308}.\\
%
\begin{figure}
\begin{center}
 \includegraphics[width=0.75\linewidth]{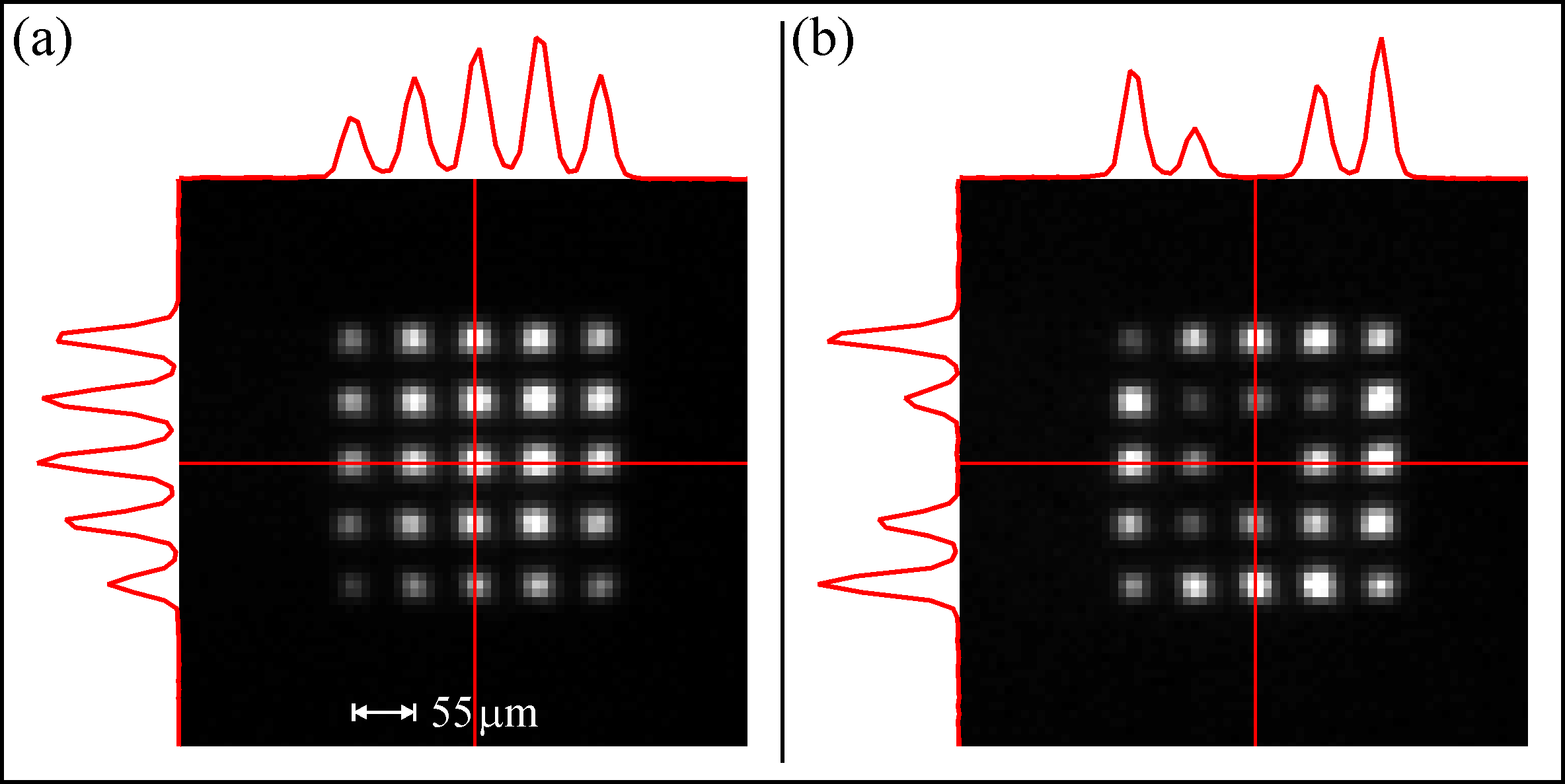}
 \end{center}
\caption{
(color online). A spatial light modulator enables the control of the trapping light power illuminating each microlens. (a) A $5\times 5$ trap pattern is implemented with full laser power at each of the 25 sites. (b) The SLM is used to change the trap depth in a reconfigurable pattern: no trapping light at the centre trap; reduced trapping light at the first ring of 8 traps; full trapping light at the outer ring of 16 traps. Scaled profiles of the detected atom fluorescence are displayed on top and at left. The reduced trap depth also reduces the number of trapped atoms and the detected fluorescence. Images are averaged 20 times.}
\label{fig:contrast}
\end{figure}
%
We extend this previous work for the experiments presented in Fig.\ \ref{fig:contrast} by adding the capability to adjust the trap depth within an 8-bit dynamic range in a site-selective fashion. Figure \ref{fig:contrast}(a) shows a fluorescence image of atom ensembles trapped in an array of microtraps with a pitch of $\unit{55}{\micro\meter}$. The spatial light modulator is used to implement a trap pattern of $5\times 5$ microtraps in a quadratic grid. Here, the sections of the SLM corresponding to the 25 microlenses creating the loaded traps are operated in full transmission, whereas all other sites of the register are switched off. The signal profiles of the central row and column exhibit a higher fluorescence level for traps located in the centre of the trap array which is caused by a larger number of stored atoms. This is due to the higher trap depth at the centre resulting from the gaussian intensity profile of the incident trapping laser beam with finite size, but also reflects the loading characteristics originating from the position and size of the MOT and the optical molasses. For the configuration displayed in Fig.\ \ref{fig:contrast}(b) we have used the SLM to modify the depths of the 25 traps in a predefined fashion with three fixed levels of light attenuation. Since a decrease in trap depth directly impacts the number of atoms loaded into each site these levels are reflected by the atomic fluorescence: the centre trap is switched off completely and no fluorescence light is detected at its position; going outwards from the centre, in the first ring of 8 traps, we have reduced the light power in such a fashion that the number of loaded atoms is reduced to about $\unit{30}{\%}$ compared to the respective traps in Fig.\ \ref{fig:contrast}(a), and finally the outer ring of 16 traps is illuminated with the same laser power as in Fig.\ \ref{fig:contrast}(a) and exhibits an accordant fluorescence level.
This proves that the implementation of our SLM-based illumination control of each lenslet can be used in an analog fashion going beyond a mere on/off switching function.
\section{Combined operation of site-selective transport and splitting of atom ensembles}
\label{sec:combined}
%
\begin{figure}
\begin{center}
 \includegraphics[width=0.75\linewidth]{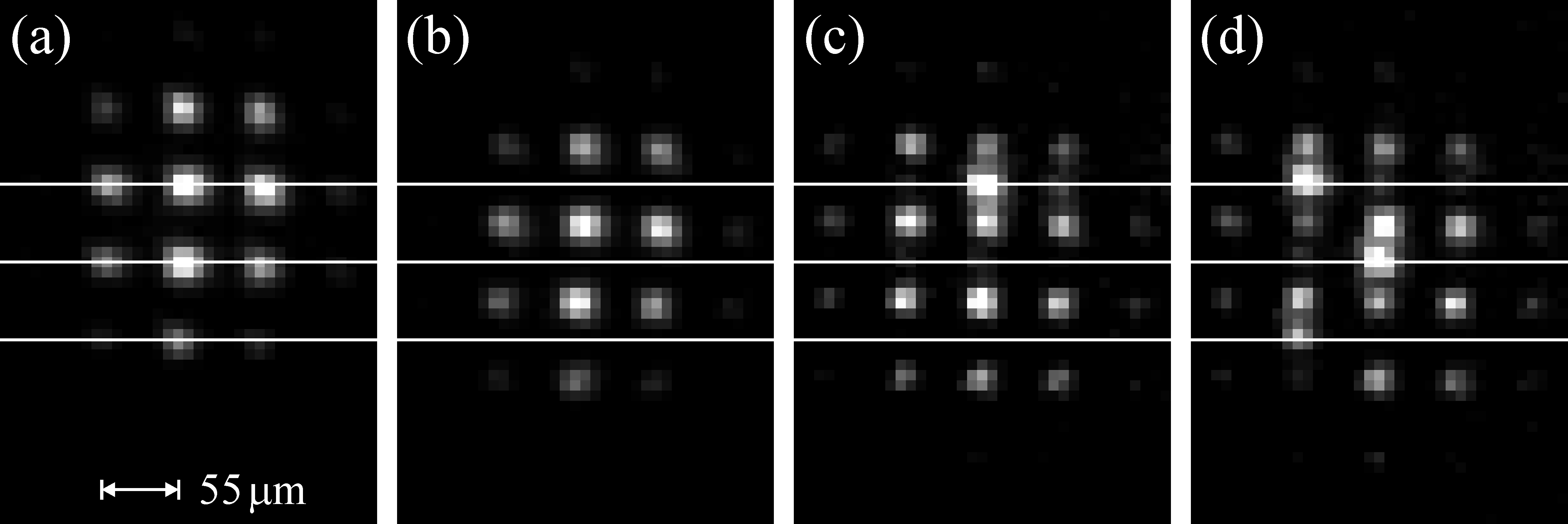}
 \end{center}
\caption{
Fluorescence images of atom ensembles stored, reloaded, and transported in a configuration of two dipole trap arrays with control of trap depth and position. (a) All traps of the SLM-controlled array A1 are switched on. Atoms are loaded and transferred to the superimposed moveable trap array A2 by decreasing the laser power in A1 and increasing it in A2 correspondingly. (b) The reloaded atom ensembles are shifted downwards by half of the trap separation ($\unit{27.5}{\micro\meter}$). (c) Site-selective splitting of atom ensembles: a single trap of register A1 is kept on while the moveable trap array is shifted downwards from the superimposed position, resulting in a splitting of the atom ensemble at the predefined register site. (d) The pattern is reconfigured to split the ensembles at three register sites. Images are averaged 20 times. 
}
\label{fig:split}
\end{figure}
%
Extending the previously reported separate implementations of SLM-con\-trol\-led site-selective addressing and atom transport in a global operation, we realized a combined system which incorporates both features (see experimental setup of Fig.\ \ref{fig:setup}). This is achieved by interleaving two separate dipole trap arrays of $\unit{55}{\micro\meter}$ pitch (created by two microlens registers at positions A1 and A2), where we have implemented single-site addressing in one and transport in the other array: we use the SLM in the beam path of A1 and the galvo mirror for position control in the beam path of A2. Both registers can be loaded with atoms and the stored atoms can be reloaded between the two trap arrays. In Fig.\ \ref{fig:split}(a) the SLM-controlled microtrap array A1 is loaded and the atom ensembles are transferred subsequently to the moveable array A2 by decreasing the trap depth of the initial array to zero and increasing the trap depth of the superimposed array correspondingly within $\unit{10}{\milli\second}$. The measured transfer efficiency is $\unit{85}{\%}$. We attribute the deviation from perfect transfer to a slight misalignment of the two trap arrays which can be avoided in an optimized setup \cite{PhysRevLett.105.170502}. Once the atoms are transferred, they are transported as shown in Fig.\ \ref{fig:split}(b) over a distance of $\unit{27.5}{\micro\meter}$ equalling one half of the trap separation.\\
The ability to dynamically superimpose and separate trap arrays whose depth and patterns can be controlled in a site-specific fashion allows us to split atom ensembles at predefined register sites. We demonstrate the splitting at a single site and at three sites in parallel in an array with $n\ge 16$ occupied dipole traps in Fig.\ \ref{fig:split}(c,d).
%
\begin{figure}
\begin{center}
 \includegraphics[width=0.75\linewidth]{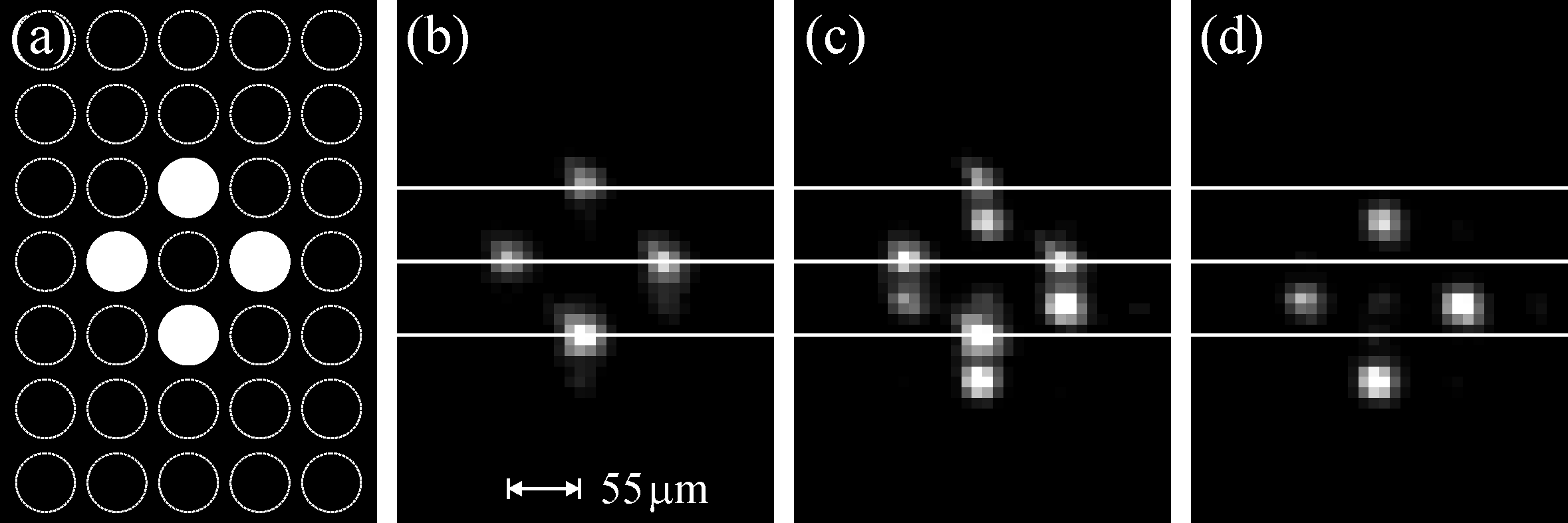}
 \end{center}
\caption{
Reconfigurable control of the splitting ratio of atom ensembles: (a) schematic view of the trap pattern (white disks) implemented by use of a SLM. Only traps indicated by the white disks are loaded with atoms. (b) A second trap register with low trap depth is superimposed and then shifted by one half of the trap separation downwards ($\unit{27.5}{\micro\meter}$). Almost all atoms remain in the unshifted register. (c) The trap depths of the two registers are approximately equal resulting in an equal splitting of the atom ensembles. (d) The trap depth of the second register is increased further and nearly all atoms leave the initial register and are transported in the second register. Images are averaged 20 times.
}
\label{fig:shifthold}
\end{figure}
%
Here, the experimental sequence is as follows: atom ensembles are stored in the position-controlled register which is superimposed with the second dipole trap array having site-selective SLM-control of the trap depth. Now, the SLM is configured to implement a desired trap pattern of comparable depth, which is a single trapping site for Fig.\ \ref{fig:split}(c) and three trapping sites for Fig.\ \ref{fig:split}(d). As the initial array with its complete set of traps starts to move most atom ensembles follow unaltered, whereas the ones at the selected sites split into two parts with atoms partially remaining at their initial position. The splitting ratio in this process is defined by the ratio of trap depths, therefore a control of the trapping laser power at each site also enables a control of the splitting ratios.\\
Figures \ref{fig:shifthold} and \ref{fig:splitratio} present a detailed analysis of this process. A trap pattern of 4 traps, as indicated by the bright circles in Fig.\ \ref{fig:shifthold}(a), is generated by the SLM-controlled array A1 and is loaded with several tens of atoms. A second array of about 16 traps, generated by the movable array A2, is superimposed after the loading sequence and then transported by one half of a trap separation downwards in Fig.\ \ref{fig:shifthold}(b-d). We have performed this experiment for different trap depths of the moveable array, while the power in the fixed array has been held constant at a value giving about equal splitting of the atom numbers in Fig.\ \ref{fig:shifthold}(c). In the case of low trap depth of the movable register, only a minor fraction of the stored atoms follows in A2 (Fig.\ \ref{fig:shifthold}(b)). For balanced trap depths, the ensembles are split into equal parts (Fig.\ \ref{fig:shifthold}(c)) and increasing the trap depth further causes the major part of the atoms to be transported in the movable array (Fig.\ \ref{fig:shifthold}(d)). The measured ratio of the number of shifted vs. unshifted atoms averaged over the four pairs of traps as a function of the depth of the shifted traps is given in Fig.\ \ref{fig:splitratio}. By changing the depth of the shifted traps, the splitting ratio can be varied in the range of 0.2 to 11, demonstrating the capability of selecting splitting ratios over almost two orders of magnitude. Turning off the power in the shifted traps completely, causes $\unit{100}{\%}$ of the atoms to remain in the unshifted traps (splitting ratio $= 0$), while the fraction of shifted atoms can be further increased to a splitting ratio $> 11$ by decreasing the power in the unshifted traps during the splitting sequence.
\section{Continuous supply of single-atom qubits}
\label{sec:singleatom}
%
%
\begin{figure}
\begin{center}
 \includegraphics[width=0.75\linewidth]{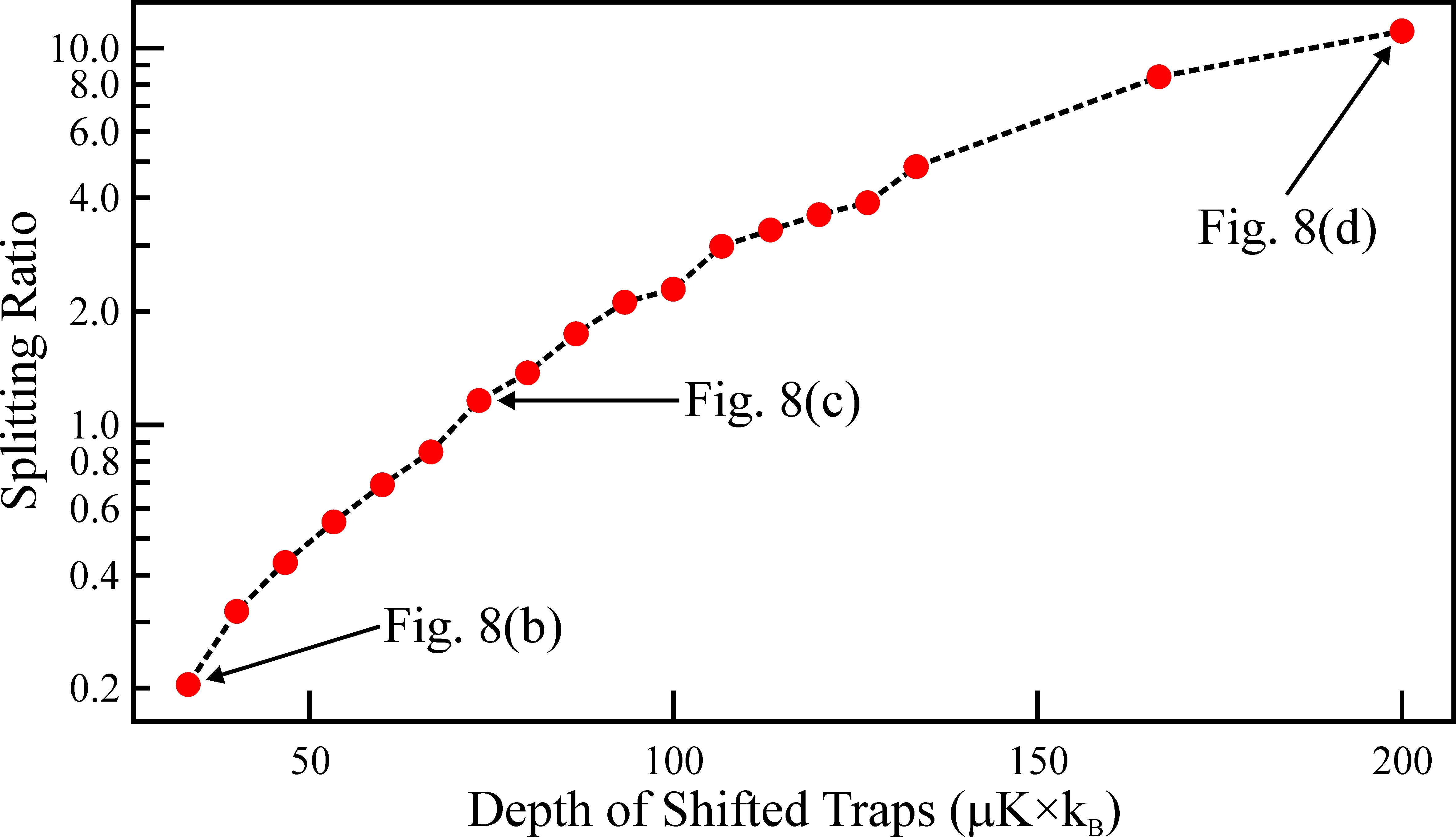}
 \end{center}
\caption{
(color online). Average ratio of the number of shifted vs. unshifted atoms (splitting ratio) as a function of the depth of the shifted traps for the four pairs of traps of Fig.\ \ref{fig:shifthold}. The respective data points of Figs. \ref{fig:shifthold}(b-d) are indicated. Data points are averaged 20 times.
}
\label{fig:splitratio}
\end{figure}
\begin{figure}
\begin{center}
 \includegraphics[width=0.75\linewidth]{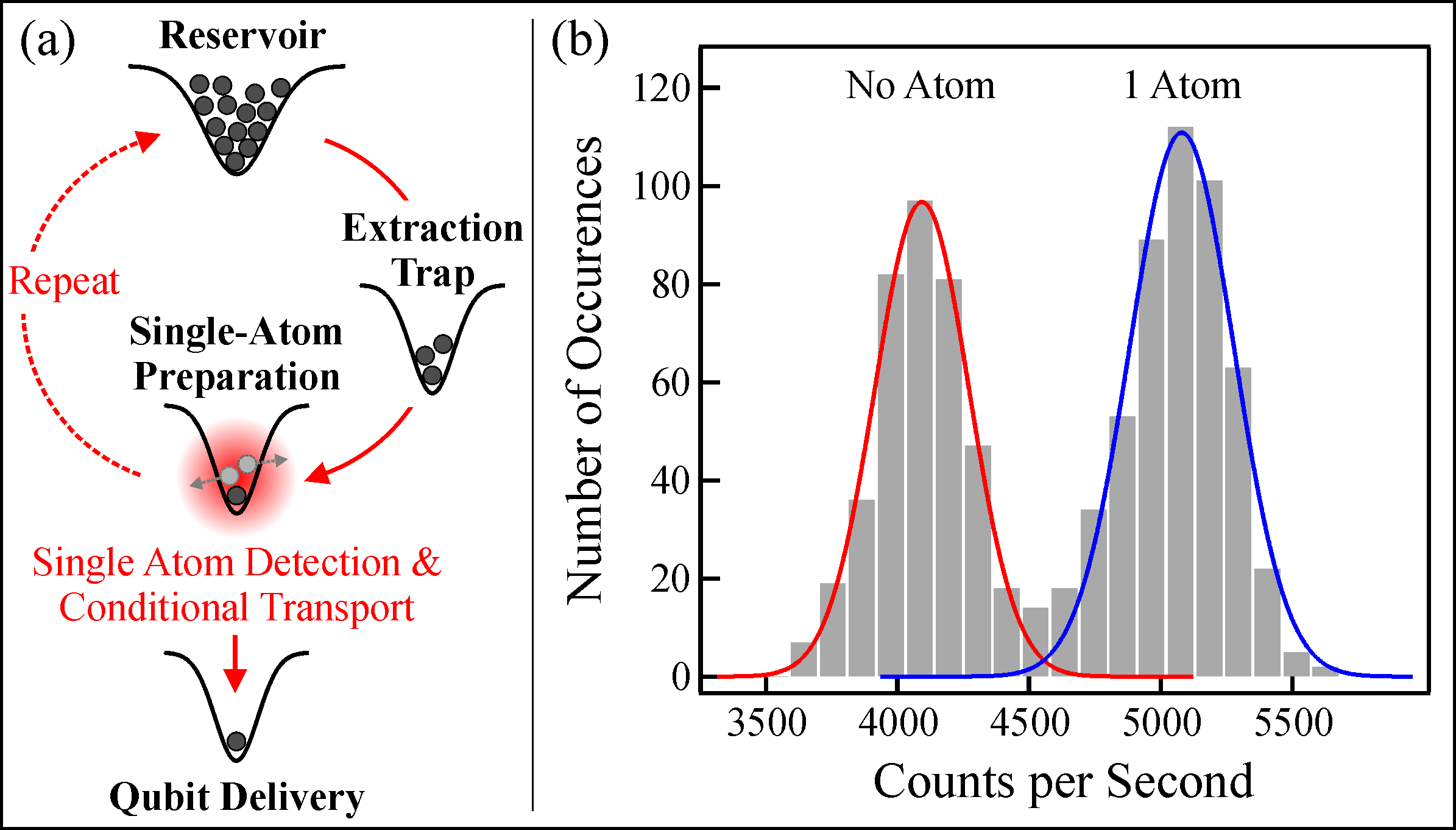}
 \end{center}
\caption{
(color online). Continuous supply of single-atom qubits.
(a) Schematic of reservoir-based single-atom qubit delivery: an optimized trapping site serves as qubit reservoir from which small atom ensembles are extracted on demand. Light assisted collisions allow for the preparation of single-atoms with sub-poissonian statistics in the extracted trap. Upon successful single-atom detection, single-atom qubits are delivered to other functional units of a quantum processor, and the extraction cycle is repeated.
(b) Histogram of recorded count rate events of a single trapping site out of a 2D register after sub-poissonian atom-number preparation. The histogram includes the photon count rate events obtained in $900$ consecutive realizations at an exposure time of $\unit{199}{\milli\second}$ (grey bars) and exhibits distinct peaks corresponding to either background light (no atoms) or the fluorescence of a single atom. Fits to the data (red and blue lines) yield the atom number probabilities for no atoms: $\unit{44.2\pm1.4}{\%}$ and 1 atom: $\unit{55.8\pm1.4}{\%}$.
}
\label{fig:singleatoms}
\end{figure}

%
The ability to split atom ensembles with selectable atom-number ratios opens an important additional path towards a scalable quantum computation architecture: in addition to the global, one-time loading scheme of the micro-trap array as implemented in our work so far, a dedicated functional unit for repeated and deterministic qubit supply can be envisioned which is separated from other functional units such as gate operation and qubit readout allowing for a continuous operation of the whole quantum circuit.\\
Figure \ref{fig:singleatoms}(a) illustrates the conceptual design of the functional unit of continuous single-atom qubit supply: one set of register sites ('Reservoir') (for clarity in the presentation only one site is shown) is optimized for the storage of a large number of cold atoms. From this, small ensembles with selectable mean atom number can be extracted on demand ('Extraction Trap').
These samples exhibiting a poissonian atom number distribution are subject to an atom-number filtering process at an intermediate trapping site. Here, site-selective addressing with near-resonant laser light is used to induce light assisted collisions which reduce the atom number to either 0 or 1 with a sub-poissonian probability distribution ('Single-Atom Preparation') as initially demonstrated in \cite{2001Natur.411.1024S}. This process can take place on a timescale of single milliseconds in optimized setups \cite{PhysRevA.85.062708} with success-rates for single-atom preparation of more than $\unit{80}{\%}$ \cite{2011NatPh...6..951G}. Subsequent fast single-atom detection can be achieved using single photon counting modules (SPCM) \cite{Bondo2006271,PhysRevLett.106.133003}. For traps with single occupancy, the single-atom qubit is further transported to the next stage in the quantum computation architecture ('Qubit Delivery'). In any case, the process is repeated by extracting the next small-number atom sample from the reservoir. This results in a quasi-continuous supply of deterministic single-atom qubits with estimated repetition rates above $\unit{100}{\second^{-1}}$ for a single unit of qubit preparation which can easily be scaled to a 2D register structure as shown above.\\
To prove this concept, we have demonstrated the controlled preparation and detection of single atoms in one of the traps of an optimized 2D register implementing a $\unit{30}{\milli\second}$-period of light assisted collisions and using an SPCM for collecting the fluorescence light scattered after this process during the detection phase. Figure \ref{fig:singleatoms}(b) shows the result: fluorescence light is integrated for $\unit{199}{\milli\second}$ and exhibits two distinct levels, as clearly visible in the histogram created from the data of $900$ consecutive experimental runs. 
The left peak corresponds to events stemming from background light only, while the right peak originates from events with a single $^{85}$Rb atom remaining in the trap. No two-atom events are observed.
The probabilities for the two possible outcomes are obtained from a fit of two gaussians to the data. The area for the no-atom signal corresponds to $\unit{44.2\pm1.4}{\%}$ probability and 1-atom events occur with a probability of $\unit{55.8\pm1.4}{\%}$. 
For a practical implementation of the single-atom qubit delivery scheme based on this data, a threshold level for sufficiently unambiguous single-atom detection could be set at e.g.\ $\unit{4833}{counts \per\second}$. Defining every event with a count rate above this level as '1 atom', corresponds to a probability of $\unit{50.0}{\%}$ for single-atom events and to a probability of $\unit{0.001}{\%}$ only of a no-atom event erroneously interpreted as a 1-atom event, resulting in a deterministic source for delivering single-atom qubits with a fidelity of $\unit{99.999}{\%}$ in - on average - every second preparation cycle.
%
%
\section{Conclusion}
\label{sec:conc}
We have extended and combined configurations for QIP based on neutral atoms in microtrap registers created by arrays of microlenses. The results presented feature global position control provided by piezo actuators, a\-cous\-to-optical beam steering, or galvo mirrors as well as SLM-enabled site-selective control of the trap depth. Along with \cite{PhysRevA.81.060308,PhysRevLett.105.170502} this demonstrates the successful combination of several essential ingredients for the implementation of a complex quantum computation architecture in a single setup, and new options for atom transport with perspective improvements in speed.\\
The application of an acousto-optical deflector allowed us to demonstrate an optical configuration significantly faster than a galvo scanner and the use of a piezo positioning system allowed us to transport $^{85}$Rb atoms stored at the sites of a $\unit{30}{\micro\meter}$-pitch register over a distance of one trap separation, which is already sufficient to connect neighbouring traps. In addition, the aberration free piezo-shift technique promotes itself for prospect 2D implementations with improved travel ran\-ges, where qubits at hundreds of trapping sites can be connected. This architecture is complemented by the control of the light addressing each microlens that allows us to implement reconfigurable trap patterns and trap depths in a site-selective fashion. Furthermore, we have demonstrated the splitting of atom ensembles at predefined register sites with adjustable splitting ratios, a technique that has prospect applications in complex architectures for QIP, such as in the integration of a qubit reservoir and opens new possibilities in the investigation of ultracold quantum gases in microtrap arrays.\\
We expect further advances in our approach from several additional options such as the capability to further down-scale the trapping geometries by optical demagnification, the application of advanced methods for coherent quantum-state transport \cite{Eckert2006264}, the incorporation of techniques for parallelized preparation and detection of single atoms in 2D arrays and the reconfigurable manipulation of the quantum states of the stored qubits in one- and two-qubit gate operations based on Rydberg blockade \cite{PhysRevLett.104.010502,PhysRevLett.104.010503}. 
\begin{acknowledgments}
We acknowledge financial support from the Deutsche Forschungsgemeinschaft (DFG), from the European Commission (Integrated Projects ACQUIRE, ACQP, and SCALA), from NIST (Grant No. 60NANB5D120), and from the DAAD (Contract No. 0804149).
\end{acknowledgments}

\bibliographystyle{apsrev4-1}
\bibliography{NJP_Schlosser}




\end{document}